# Efficient Tabling Mechanisms for Transaction Logic Programs

Paul Fodor
Computer Science Department
Stony Brook University
Stony Brook, NY 11794

**Abstract**
In this paper we present efficient evaluation algorithms for the Horn Transaction Logic (a generalization of the regular Horn logic programs with state updates). We present two complementary methods for optimizing the implementation of Transaction Logic. The first method is based on tabling (memoing for logic programs) and we modified the proof theory to table calls and answers on states (practically, equivalent to dynamic programming). The call-answer table is indexed on the call and a signature of the state in which the call was made. The answer columns contain the answer unification and a signature of the state after the call was executed. The states are signed efficiently using a technique based on tries and counting. The second method is based on incremental evaluation and it applies when the data oracle contains derived relations. The deletions and insertions (executed in the transaction oracle) change the state of the database. Using the heuristic of inertia (only a part of the state changes in response to elementary updates), most of the time it is cheaper to compute only the changes in the state than to recompute the entire state from scratch. The two methods are complementary by the fact that the first method optimizes the evaluation when a call is repeated in the same state, and the second method optimizes the evaluation of a new state when a call-state pair is not found by the tabling mechanism (i.e. the first method). The proof theory of Transaction Logic with the application of tabling and incremental evaluation is sound and complete with respect to its model theory. The application of these algorithms promises great improvements in the applications of transaction logic dealing with state-changing systems (e.g. systems involving financial transactions), dynamic constraints on transaction execution (e.g. workflow modeling and verification) and applications of artificial intelligence planning (e.g. discovery and contracting of Semantic Web Services).

## 1 Introduction

Transaction Logic is a logic designed for programming state-changing actions, executing them, and reasoning about their effects developed by Anthony Bonner of University of Toronto and Michael Kifer of Stony Brook University [BonnerKiferReport95]. Transaction Logic (TR) was used to perform various practical analyses which need sate-updates and transactions, such as: workflows modelling [BonnerKiferReport95, DavKif98, DavKifRam04], program analysis and verification [BonnerKiferReport95], AI planning [BonnerKiferReport95]. It has a general model theory and a sound and complete proof theory with respect to its model theory. The transaction logic implementations concentrate on performing the analysis correctly for the general case, but are not very efficient. This paper concentrates on developing efficient algorithms for the Horn Transaction Logic (a generalization of the regular Horn programs with updates and transaction) based on two complementary methods: tabling and incremental evaluation.
In the following sections we will introduce the reader to transaction logic and the optimization methods of the transaction logic implementation.

## 2 Transaction Logic

Transaction Logic is a sound and complete logical formalism for state changing domains. It is a general theory that contains no specification of the nature of the states being updated (e.g. relational databases, logic programs, non-logical theories) or of the nature of the updates (e.g. tuple insertion/deletion, relational SQL-style bulk updates, non-logical state changes done by an algorithm). Instead, it uses data oracles to solve queries on the states and transaction oracles to specify the updates and the effect of the updates on the states. If the data and transaction oracles are specified, then Transaction Logic can be used to reason about the effects of the actions. The Transaction Logic isolates the details of state semantics from the rest through data oracles: $\mathcal{O}^d$: States → Sets of First-order, where the formulas $\mathcal{O}^d(s)$ tells the logic what is true at state s. At the same time the Transaction Logic isolates the transitions between states (elementary updates) through transition oracles: $\mathcal{O}^t$: States × States → SetsofElementaryUpdatesOfGroundAtoms. An elementary updates of a ground atom b ∈ $\mathcal{O}^t(D_1, D_2)$ means, executing b causes state transition from state $D_1$ to $D_2$
An example of this semantics is deductive databases, where the states are relational databases and the state transitions can be of only these kinds:

    Insert: p.ins($t_1$, …, $t_n$) ∈ $\mathcal{O}^t(D_1, D_2)$ iff ∈ $D_2 = D_1$ ∪ {p($t_1$, …, $t_n$)}

    Delete: p.del($t_1$, …, $t_n$) ∈ $\mathcal{O}^t(D_1, D_2)$ iff ∈ $D_2 = D_1$ - {p($t_1$, …, $t_n$)}

The syntax of transaction logic defines transaction formulas built of transaction goals (atoms) and the connectives from Figure 1. A transaction goal can be either an asignment of the data oracle function or of the transaction oracle function. In the case of deductive databases a transaction goal can be either a predicate atom (equivalent to the atom in predicate calculus) or an elementary update (e.g. insert or delete) of a predicate atom.

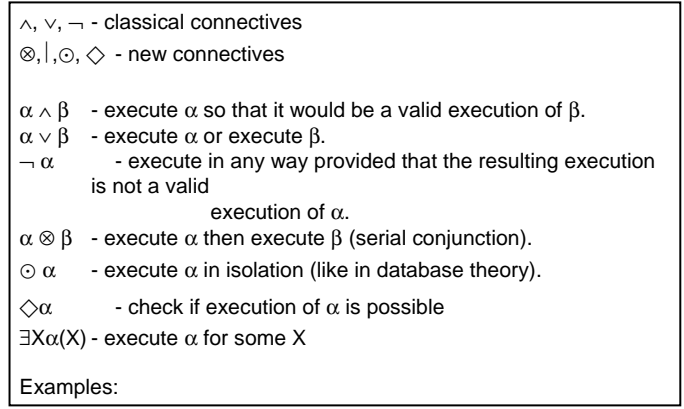

Figure 1. Transaction Logic Syntax

The modal theory semantics of transaction logic is formalized around the concept of paths. Any formula in Transaction Logic is a transaction and has truth values over execution paths (and not over states). A path is a sequence of states. A transaction φ is true on a path $\pi=\langle s_1, s_2, s_3, \ldots, s_{n-2}, s_{n-1}, s_n\rangle$ means that φ can execute at state $s_1$, changing it to state $s_2$, ..., to state $s_n$, terminating at state $s_n$. This means that evaluating the truth value over a path is equivalent to the execution over that path. For the Concurrent Transaction Logic (Transaction Logic extended with the concurrent conjunction operator "|", where "α|β" means execute α and β in parallel) the semantics is extended to multi-paths, i.e. paths with "pauses", other transactions can execute during those pauses and so, transactions are interleaved). Queries are transactions that execute over paths ⟨s⟩ (do not change state) and when execution is restricted to paths of length 1, Transaction Logic reduces to classical logic. Intuitively, the transaction logic checks if a formula φ can be executed on a path formed of database states (we say executes φ along the path π as it proves φ), as in Figure 2.

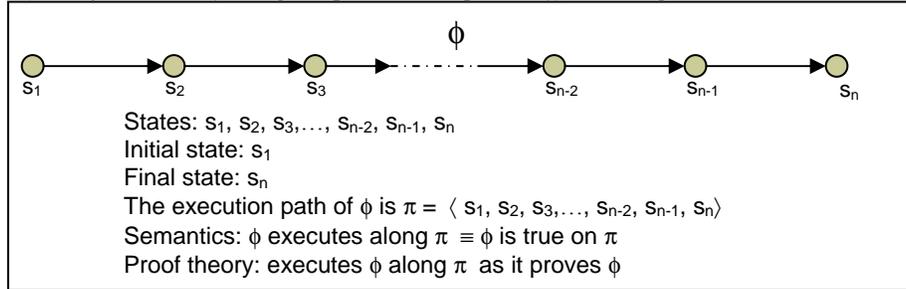

Figure 2. Transaction Logic Semantics

A path structures is a function defined on the set of all possible paths that assigns first-order semantics to paths:

$M$: Paths → FirstOrderSemanticStructures

If D is a state, φ is a first order formula, and $\mathcal{O}^d(D) \vDash^c \phi$ ($\vDash^c$ denoting classical logical entailment), then $M(\langle D\rangle) \vDash^c \phi$ (this property is denoted data oracle compliance). If $\mathcal{O}^t(D_1, D_2) \vDash^c \psi$ then $M(\langle D_1, D_2\rangle) \vDash^c \psi$ (i.e. transition oracle compliance). The base case is:

$M, \pi \vDash p(t_1, t_2, \ldots, t_n)$ iff $M, \pi \vDash^c p(t_1, t_2, \ldots, t_n)$ for any atomic formula (query or transaction invocation) $p(t_1, t_2, \ldots, t_n)$

All the operators are defined with respect to the path structure, such as:

Negation: $M, \pi \vDash \neg \phi$ iff not($M, \pi \vDash^c \phi$) (cannot execute φ along the path π)

"Classical" conjunction: $M, \pi \vDash \phi \wedge \psi$ iff $M, \pi \vDash^c \phi$ and $M, \pi \vDash^c \psi$ (can execute φ and ψ along the same path).

Serial conjunction: $M, \pi \vDash \phi \otimes \psi$ iff $M, \pi_1 \vDash^c \phi$ and $M, \pi_2 \vDash^c \psi$ for some paths $\pi_1$ and $\pi_2$ such that $\pi = \pi_1 \circ \pi_2$ (do φ then ψ).

Possibility: $M, \langle s_1\rangle \vDash \Diamond \phi$ iff there is a path $\pi = \langle s_1, \ldots, s_n\rangle$ such that $M, \pi \vDash \phi$ ($\Diamond \phi$ is always a query – true at states, even if φ executes over a sequence of states).

An example of path structures of transaction logic can be seen in Figure 3.

> $\phi_1$ = nodeA.del $\otimes$ nodeB.ins $\otimes$ nodeC $\otimes$ nodeD.ins
> Initial state: {nodeA,nodeC}
> $\phi_1$ is true over path
> $\quad\quad \pi = \langle$ {nodeA,nodeC}, {nodeC}, {nodeB, nodeC}, {nodeB, nodeC,nodeD} $\rangle$
> Let $M$ be a path structure
> By the definition of the oracles and path structures:
> $\quad\quad \mathcal{O}^t(\{\text{nodeA, nodeC}\}, \{\text{nodeC}\}) \vDash$ nodeA.del, hence
> $\quad\quad\quad\quad M, \langle\{\text{nodeA,nodeC}\}, \{\text{nodeC}\}\rangle \vDash$ nodeA.del
> $\quad\quad \mathcal{O}^t(\{\text{nodeC}\}, \{b, \text{nodeC}\}) \vDash$ nodeB.ins,
> $\quad\quad\quad\quad$ hence $M, \langle\{\text{nodeC}\}, \{\text{nodeB,nodeC}\}\rangle \vDash$ nodeB.ins
> $\quad\quad \mathcal{O}^d(\{\text{nodeB, nodeC}\}) \vDash$ nodeC, hence $M, \langle\{\text{nodeB,nodeC}\}\rangle \vDash$ nodeC
> $\quad\quad \mathcal{O}^t(\{\text{nodeB, nodeC}\}, \{\text{nodeB,nodeC,nodeD}\}) \vDash$ nodeD.ins, hence
> $\quad\quad\quad\quad M, \langle\{\text{nodeB,nodeC}\}, \{\text{nodeB,nodeC,nodeD}\}\rangle \vDash$ nodeD.ins
> By definition of $\otimes$ implies that then $M,\pi \vDash \phi_1$

Figure 3. Simple Example of Transaction Logic

A *transaction program* is a set of transaction formulas P. M is a model of P iff $M,\pi \vDash \phi$ for every path $\pi$ and every $\phi \in$ P. If $\phi$ is a transaction formula, and $D_0, D_1,\ldots, D_n$ is a sequence of database state identifiers, then execution entailment is a statement of the form:

P, $D_0, D_1,\ldots, D_n \vDash \phi$ and it means: M, $D_0, D_1,\ldots, D_n \vDash \phi$ for every model M of P.

The proof theory of transaction logic is sound and complete with respect to the above model theory. One possible such inference system is $\mathcal{F}^I$ [BonnerKiferReport95] (see Figure 4).

> Axioms: P, D --- $\vdash$ () .
> Inference rules:
> $\quad$ Applying transaction definitions: let a $\leftarrow \phi \in$ P, a and b unify with mgu $\sigma$
>
> $\quad\quad\quad$ P, D --- $\vdash (\exists)(\phi \otimes$ rest$)\sigma$
> $\quad\quad\quad$ ─────────────────────
> $\quad\quad\quad$ P, D --- $\vdash (\exists) (b \otimes$ rest)
>
> $\quad$ Querying the database: if b$\sigma$ and rest $\sigma$ share no variables, and $\mathcal{O}^d(D) \vDash^c (\exists)b\,\sigma$,
>
> $\quad\quad\quad$ P, D --- $\vdash \quad (\exists)$ rest$\sigma$
> $\quad\quad\quad$ ─────────────────────
> $\quad\quad\quad$ P, D --- $\vdash (\exists) (b \otimes$ rest)
>
> $\quad$ Performing elementary updates: if b$\sigma$ and rest $\sigma$ share no variables, and $\mathcal{O}^t(D_1,D_2)\vDash^c (\exists)b\sigma$, then
>
> $\quad\quad\quad$ P, $D_2$ --- $\vdash \quad (\exists)$ rest$\sigma$
> $\quad\quad\quad$ ─────────────────────
> $\quad\quad\quad$ P, $D_1$ --- $\vdash (\exists) (b \otimes$ rest)
>
> **Soundness of** $\mathcal{F}^I$: let $\phi$ be a serial goal, then
> $\quad$ If P, D --- $\vdash (\exists) \phi$ then P, D --- $\vDash (\exists) \phi$
> **Completeness of** $\mathcal{F}^I$: let $\phi$ be a serial goal, then
> $\quad$ If P, D --- $\vDash (\exists) \phi$ then P, D --- $\vdash (\exists) \phi$

Figure 4. Simple Example of Transaction Logic

Sequential Horn Transaction Logic is a subset of transaction logic where all rules in the program have the format: "atom$\leftarrow$Goal", where Goal can be an atomic formula ($\phi_1 \otimes \ldots \otimes \phi_k$), where each $\phi_i$ is a concurrent serial goal; or $\odot\phi$, where $\phi$ is a concurrent serial goal. An example of a Sequential Horn Transaction Logic is a financial transaction like the one in Figure 5.

> transfer(Amt,Acct1,Acct2) $\leftarrow$ withdraw(Amt,Acct1) | deposit(Amt,Acct2)
> withdraw(Amt,Acct) $\leftarrow \odot$(balance(Acct, Bal) $\otimes$ Bal $\geq$ Amt
> $\quad\quad\quad\quad\quad \otimes$ changeBallance(Acct, Bal, Bal - Amt))
> deposit(Amt,Acct) $\leftarrow \odot$(balance(Acct, Bal) $\otimes$ changeBallance(Acct, Bal, Bal+Amt))
> changeBallance(Acct,Bal1,Bal2) $\leftarrow$ balance.del(Acct,Bal1)$\otimes$balance.ins(Acct,Bal2)
> Query: $\quad$ ?- transfer(Fee,Client,Broker) $\otimes$ transfer(Cost,Client,Seller)

Figure 5. Financial transaction example

The applications of Transaction Logic are diversified [BonnerKiferReport95]: consistency maintainance, view updates, heterogeneous databases, systems with state and dynamic constraints on the execution of the transaction, AI planning.

## 3 Tabling for Transaction Logic

Tabling or memoing is a technique to memoize the results of computations to avoid repeated sub-computations. The technique was applied in logic programming [TamSat86], [War92] by recording the goals in the calls and provable instances (answers) in a call-answer table. On encountering a goal G, if a variant of G is present in call table: G is resolved with the associated answers. If a variant of G is not present in call table: G is entered in the call column of the calls-answers table and G is resolved with program clauses to generate answers. Each answer is entered in the associated answer column if its not there.

An example of the application of OLDT refutation method for the logic program from Figure 6 and the goal: "reach(a,X)" is in Figure 7. The call-answer table is depicted in Table 1.

```
edge(a,b).
%%%  r/2: left recursive transitive closure
reach(X,Y) :- edge(X,Y).
reach(X,Y) :- reach(X,Z), edge(Z,Y).
```

Figure 6. Logic programming for computing the transitive closure

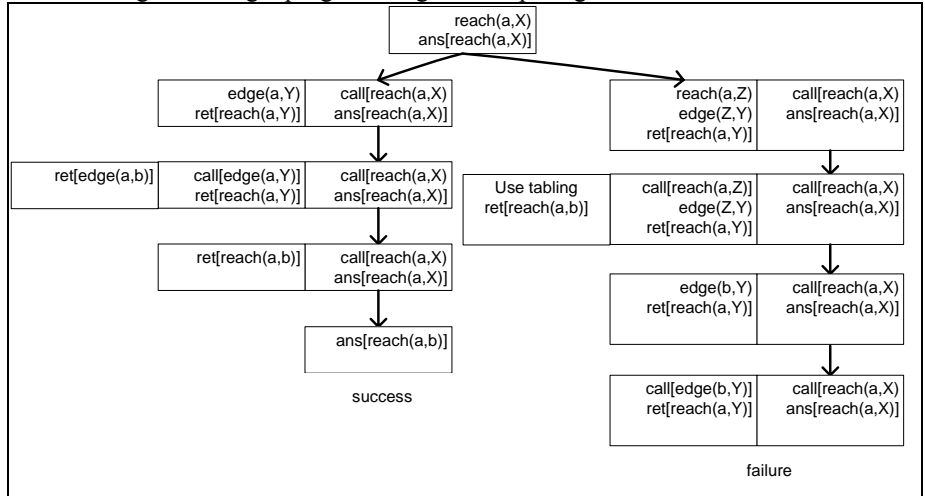

Figure 7. OLDT refutation in Logic Programming

| Call | Answer Unification |
|---|---|
| reach(a,X) | [X/b] |
| edge(a,Y) | [Y/b] |
| edge(b,Y) | fail |

Table 1. Call-answer table

Memoing for Transaction Logic promises an optimization of the execution since it produced very good optimizations for logic programming. Tabling becomes more complex for Transaction Logic because of state and path semantics, that is: it is not the same state of the database when the entries in the call table are called. We present here a method for tabling for Transaction Logic that is a solution for the general theory of Transaction Logic (and not only of Horn Transaction Logic). The idea is to modify the tabling mechanism to table call and the state in which the call is made, and to produce answer unifications and final states after the execution of the call took place. Our new algorithm, denoted OLDT-TR refutation contains the state added on the right side of each of the nodes in OLDT refutation. Before we present the algorithm, we will first show how tabling can be applied in Transaction Logic by means of an example. Lets consider that we have the Transaction Logic program from Figure 8. Then the OLDT-TR refutation is presented in Figure 9. The call-answer table is depicted in Table 2.

```
Data oracle:
edge(1,2).

Transaction oracle:
reach(X,Y)←edge(X,Y)⊗del.edge(X,Y).
reach(X,Y)←edge(X,Z) ⊗del.edge(X,Z) ⊗reach(Z,Y).
```

Figure 8. Transaction Logic program for computing the transitive closure

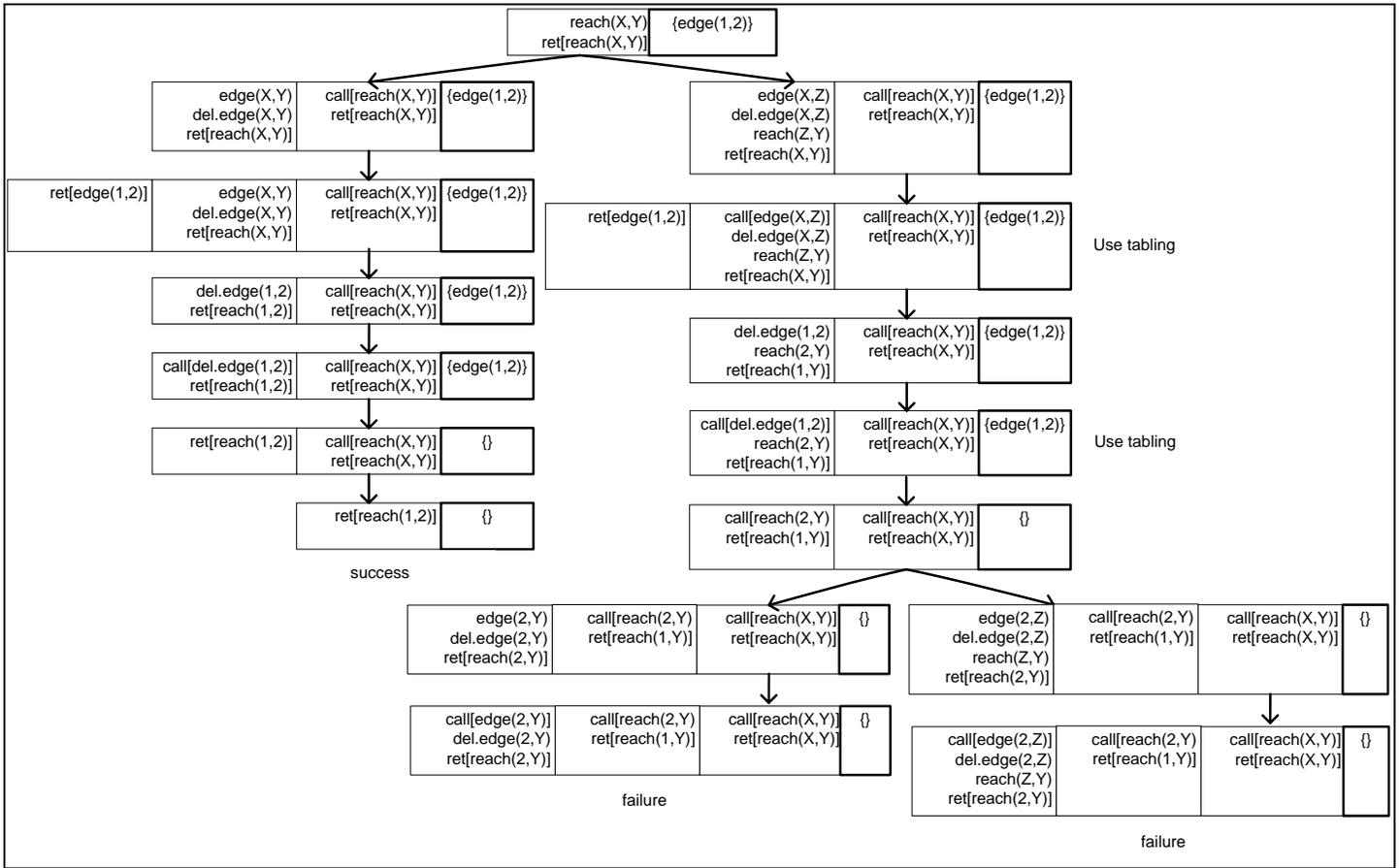

Figure 9. OLDT-TR refutation

| Call | Initial State | Answer Unification | Final State |
|---|---|---|---|
| reach(X,Y) | {edge(1,2)} | [X/1,Y/2] | {reach(1,2)} |
| edge(X,Y) | {edge(1,2)} | [X/1,Y/2] | {edge(1,2)} |
| del.edge(1,2) | {edge(1,2)} | [] | {} |
| reach(2,Y) | {} | fail | |
| edge(2,Y) | {} | fail | |

Table 2. OLDT-TR call-answer table

One drawback of this algorithm is that we have to store the current state of the database in the call-answer table, which is unacceptable. Our solution to this problem is to consider that the OLDT-TR computation started from an initial database $D_0$, so we can use the log of changes of the database, instead of the entire current state of the database (see example in Figure 10). This is only an heuristic and some exceptional cases (e.g. bulk updates) it may be cheaper to store the state than the changes to the state.

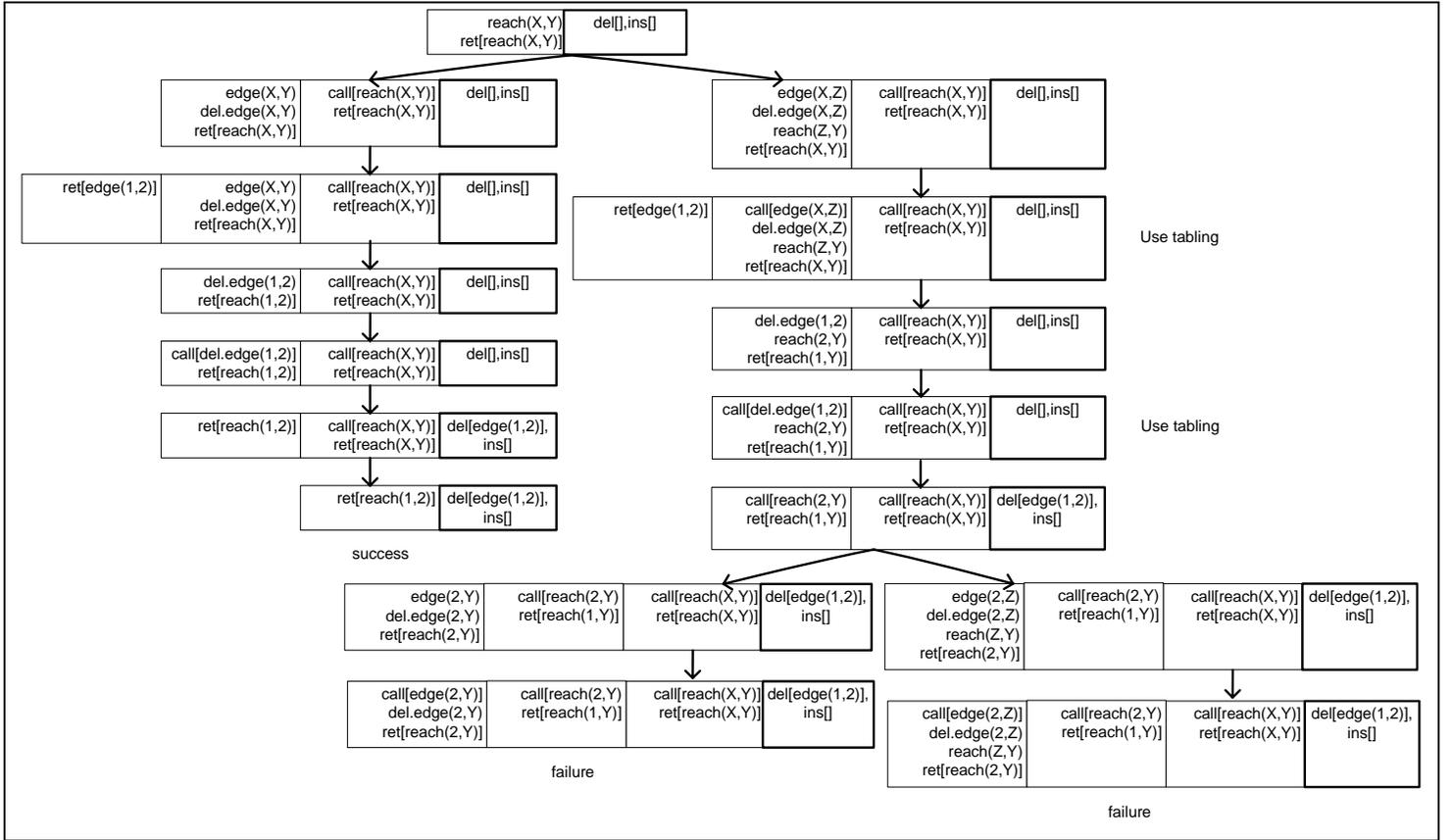

Figure 10. OLDT-TR refutation using logs instead of states

Unfortunatelly, even the update logs can become very big, so we thought of keeping only a signature of the log in the node instead of the actual log. We developed two different efficient methods to uniquely identify the state of a database. The steps for both these algorithms are: computing the signature of the rules in the data oracle, for a given state order the two lists of insert and delete signatures, eliminate reverse operations (i.e. ins.P and del.P are reverse operations - if a reverse relation is already in the log (ins.p vs. del.p) then they reverse each other's results), sign the logs with a new pair of signatures. The advantage of these solutions are: comparing two signed states has a score of O(1) and computing the new signatures can be done incrementally.

In the following paragraphs we will present the two methods for signing states efficiently: one based on tries and counting (the practical method) and a second method based on Gödel numbering.

The first method of signing states is based on two trie data structures (global rule identification trie and state log identification trie) and has the following steps (developed incrementally):
- incrementally sign all the rules that can appear in a log at the time they appear by using a trie and a counter (by incrementing the counter each time a new rule is found),
- given an exiting log of inserts and deletes and an elementary transaction operation, find and eliminate the reverse operation or insert the reverse operation in the corresponding list of inserts and deletes,
- incrementally compute the signatures of the state logs at the time they appear by using a trie of rule signatures and a counter (by incrementing the counter each time a new state is found).

Let's consider the following example: we have an initial database and we insert the following facts: e(1,2), e(1,3), e(2,3) and r(V1,V2)←e(V1,V3)⊗r(V3,V2) and then we delete e(1,3). The execution path of the database will go through the execution path described by the log of states from Table 3. Our algorithm incrementally computes the RTRoot and STRoot tries, and RTCounter and STCounter counters (see Figure 11 and Table 4). Practically, this algorithm generates a new rule identifier if a rule was not found, or generates a new state identifier if a state was not found in the previous states.

| State Logs |
|---:|
| ins[e(1,2)],del[] |
| ins[e(1,2), e(1,3)],del[] |
| ins[e(1,2) , e(1,3) , e(2,3)],del[] |
| ins[e(1,2) , e(1,3) , e(2,3), r(V1,V2)←e(V1,V3)⊗r(V3,V2)],del[] |
| ins[e(1,2), e(2,3), r(V1,V2)←e(V1,V3)⊗r(V3,V2)],del[] |

Table 3. State Logs

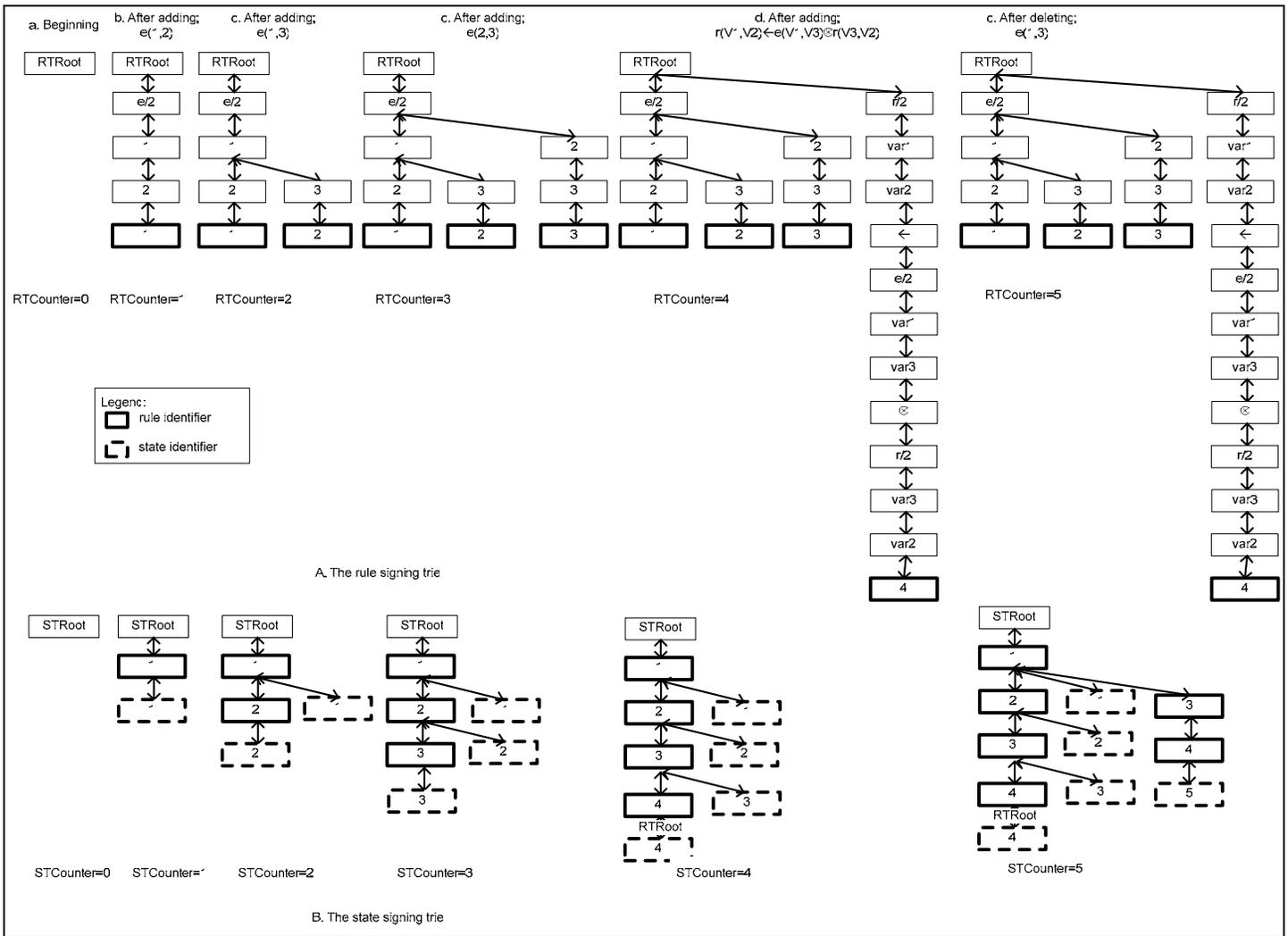

Figure 11. Rules and state logs labeling using tries

| State Log | State signature |
|---|---|
| ins[e(1,2)],del[] | 1 |
| ins[e(1,2), e(1,3)],del[] | 2 |
| ins[e(1,2), e(1,3), e(2,3)],del[] | 3 |
| ins[e(1,2), e(1,3), e(2,3), r(V1,V2)←e(V1,V3)⊗r(V3,V2)],del[] | 4 |
| ins[e(1,2), e(2,3), r(V1,V2)←e(V1,V3)⊗r(V3,V2)],del[] | 5 |

Table 4. State log identification

The second method of signing states is based on Gödel numbering and works as follows:
- incrementally sign all the rules that can appear in a log at the time they appear by using a Gödel numbering,
- given an exiting log of inserts and deletes (represented as rule signatures) and an elementary transaction operation, find and eliminate the reverse operation or insert the reverse operation in the corresponding list of inserts and deletes,
- incrementally compute the signatures of the state logs at the time they appear by using a Gödel numbering for the inserts and another Gödel numbering for the deletes.

Let's consider the following alphabet for the rule labeling of the Gödel numbering as the map of the finite set $\{(,),\leftarrow,\otimes,.,a,b,c,d,e,f,g,h,…,z,A,B,…,Z,0,1,2,3,4,5,6,7,8,9\}$ to the position in the set (i.e. "("→1, ")"→2, "←"→3,…). The Gödel numbering given a rule R=" $r_1 r_2 r_3 …r_n$", labels the rule with a natural number as follows: $2^{map(r1)} * 3^{map(r2)} * 5^{map(r3)} * … * p^{map(rn)}$. For example "e(a)" is labelled as the natural number: $2^{10}*3^1*5^2*7^6$

The second step of this algorithm consider the alphabet for the state labeling of the Gödel numbering as the map of the finite set $\{rule_1, rule_2,…, rule_n\}$ to the labels computed in the first step: $\{ruleLabel_1, ruleLabel_2,…, ruleLabel_n\}$. The Gödel numbering given a ordered insert log S=" $rl_1 rl_2 rl_3 …rl_n$", labels the rule with a natural number as follows: $2^{rl1} * 3^{rl2} * 5^{rl3} * … * p^{rln}$. For example "ins[e(a)]" is labelled as the natural number: $2^{\wedge}(2^{10}*3^1*5^2*7^6)$.

Unfortunatelly, since the numbers created by this second method are huge, this method has only theoretical value and is impossible to be applied in practice.

Finally, we present the OLDT-TR $\mathcal{F}^I$ inference system

Data structures:
  rule identification trie
  state identification trie
  call-answer table of 4 columns: call,callInitialStateId,answerUnification,returnStateId

Axiom: If the query is P, D --- ⊢ () return true
Inference steps:
  1. Tabling definitions: if b is a transaction goal (either query or elementary update) and it unifies with an entry c in the call-answer table (call column (c,D)) with mgu σ

  $$\frac{P, D --- \vdash (\exists)(c \otimes rest)\sigma}{P, D --- \vdash (\exists) (b \otimes rest)}$$

  else save (b,D) in the call-answer table (in the call column) with empty entry for the answer column and continue with step 2
  2. Applying transaction definitions: if a ← φ ∈ P, a and b unify with mgu σ, then

  $$\frac{P, D --- \vdash (\exists)(\phi \otimes rest)\sigma}{P, D --- \vdash (\exists) (b \otimes rest)}$$

  3. Querying the database: if bσ and rest σ share no variables, and Od(D) ⊨c (∃)b σ, then

  $$\frac{P, D --- \vdash \quad (\exists) rest\sigma}{P, D --- \vdash (\exists) (b \otimes rest)}$$

  4. Performing elementary updates: if bσ and rest σ share no variables, and Ot(D1,D2) ⊨c (∃)bσ, then

  $$\frac{P, D2 --- \vdash \quad (\exists) rest\sigma}{P, D1 --- \vdash (\exists) (b \otimes rest)}$$

  If the query is P, D --- ⊢ () return true and fill the answer table with (answerUnification, returnState) for all at first level of derivation

We apply the OLDT-TR $\mathcal{F}^t$ refutation to the program from Figure 8 and the goal: "reach(X,Y)" and we obtain the refutation of Figure 12, the call-answer of Table 2, where the log of the states are labelled as in Table 5.

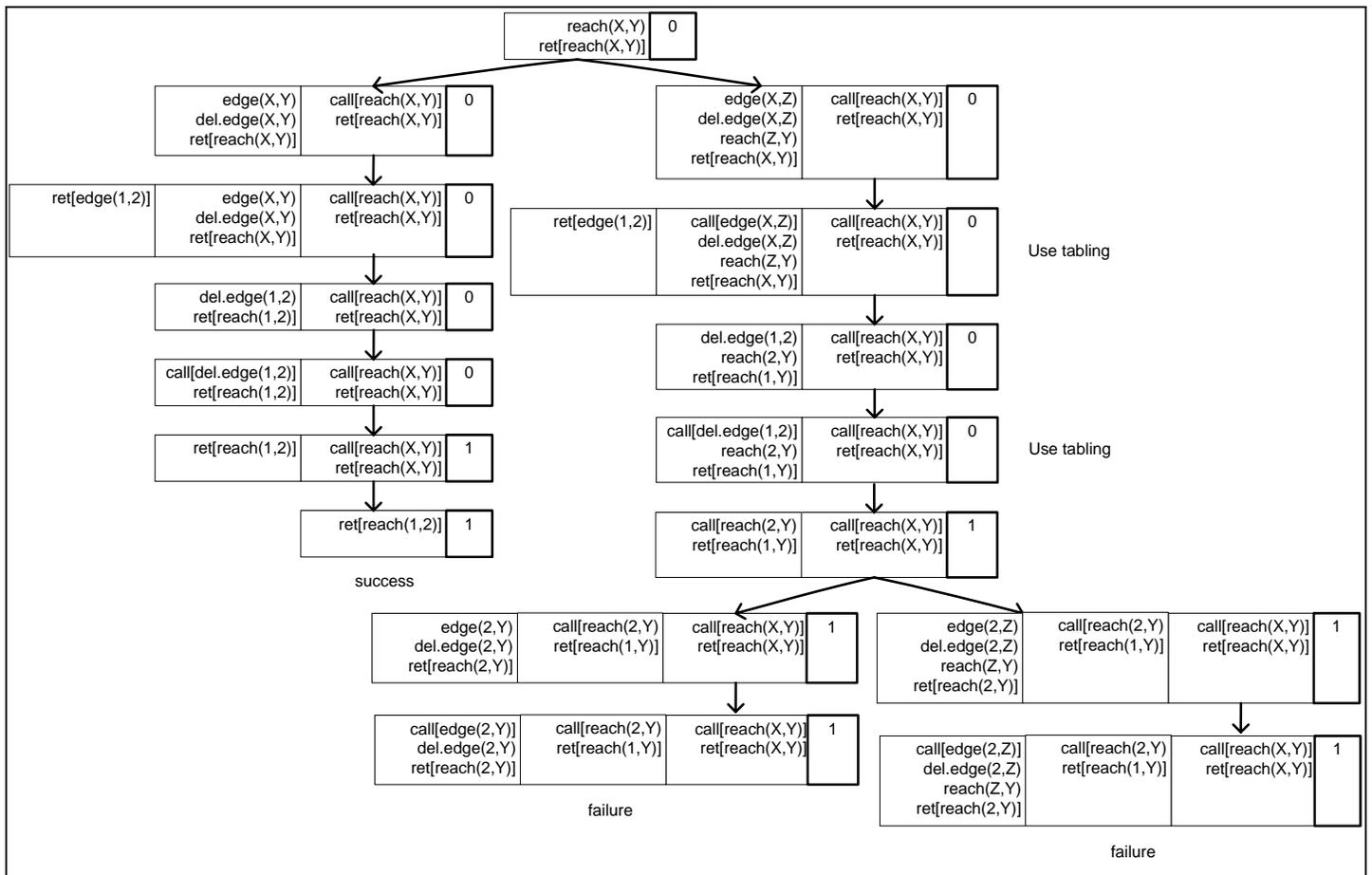

Figure 12. OLDT-TR $\mathcal{F}^t$ example using trie singning

| State Log | State signature |
|---|---|
| ins[],del[] | 0 |
| ins[edge(1,2)],del[] | 1 |

Table 5. State log identification

| Call | Initial State Signature | Answer Unification | Final State Signature |
|---|---|---|---|
| reach(X,Y) | 0 | [X/1,Y/2] | 0 |
| edge(X,Y) | 0 | [X/1,Y/2] | 0 |
| del.edge(1,2) | 0 | [] | 1 |
| reach(2,Y) | 1 | fail | - |
| edge(2,Y) | 1 | fail | - |

Table 6. OLDT-TR call-answer table

## 4 Incremental Evaluation of Tabled Transaction Logic

Tabling takes care of calls that were executed before in the same state. Unfortunatelly, an important part of transaction logic was not optimized by tabling and that is when a call was not encounted before. To optimize the data oracle for a call that was not encounted before, we use an idea generated from maintenance of materialized views in databases (more precisely, the Delete-Rederive algorithm (DRed) from [GupMum93]). Elementary updates (e.g. deletions and insertions) executed by the transition oracle change the state of the database and re-computing the state from scratch is too wasteful in most cases. Using the heuristic of inertia (only a part of the state changes in response to elementary updates), it is often cheaper to compute only the changes in the state than re-computing the state from scratch. Thismethod is only an heuristic, in some cases (e.g. bulk updates) it may be cheaper to re-compute a state than to compute the changes to the state. For example the program of Figure 13 is a Transaction Logic to find if a graph has a Hamiltonian path using predecessor consumption. In this program the predicate choose/2 is a derived predicate in the data oracle. It is easy to see that the derications of the choose predicate can be easily maintained using incremental analysis (add figure).

```
%%%  path/0: find Hamilton paths by consuming nodes:
path ← n(N) ⊗ del.n(N) ⊗ extend(N) ⊗ ins.n(N)
extend(N₁) ← choose(N₁,N₂) ⊗ del.n(N₂) ⊗ extend(N₂) ⊗ ins.n(N₂)
extend(N) ← empty.n
choose(N₁,N₂) ← e(N₁,N₂) ⊗ n(N₂)
```

Figure 13. Search for Hamiltonian paths by consuming nodes

Combining incremental evaluation with tabling is not trivial. If a call is not found in the table then after the execution of that call the support graph has to be stored in the table (which implies that one also has to table the support graph), otherwise the support graph has to be loaded from the table. There might be more than one solution. A first solution is to define the difference between two "logs" ({ins[e(1,2)],dels[e(1,3)]} \ {ins[e(1,5)],dels[e(2,3)]} = {ins[e(1,3),e(1,5)],dels[e(1,2),e(2,3)]}), keep the current support graph and apply the difference of logs on this support graph.

A second solution is to keep a full complete support graph, sign the support nodes and memorize for each state what is the list of its support nodes (see example below).

Lets have the rules:

r(X,Y) ← e(X,Y)
r(X,Y) ← e(X,Z) ⊗ r(Z,Y)

The Figure 14 is the support graph having the facts: e(1,2) and e(2,3).

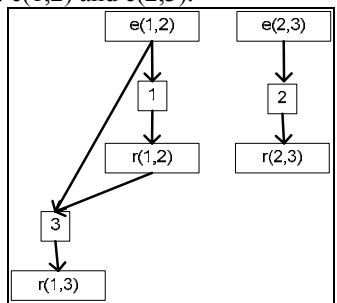

Figure 14. Support graph for the r/2 rule

The solution to determine at each step the corresponding support graph resumes to signing the support nodes (e.g. 1,2,3) and memorize for each state what is the list of its support nodes (see table 7).

| State Log | State support nodes |
|---|---|
| ins[],del[] | {1,2,3} |
| ins[],del[e(1,2)] | {2} |
| ins[],del[e(2,3)] | {1} |
| ins[],del[e(1,2), e(2,3)] | {} |

Table 7. State log support nodes in the support graph

**Related Work**

There are a set of Transaction Logic implementations: Flora2, UTorontoCTR and MEU. In Flora-2 the primitives btdelete and btinsert implement the insert and delete operators with the Transaction Logic semantics (implemented using the XSB tries, but with no tabling on the state identifier):

?R[%stack(0, ?X)] : − ?R:robot.
?R[%stack(?N, ?X)] : − ?R:robot, ?N > 0,
                     ?Y[%move(?X)], ?R[%stack(?N − 1, ?Y)].
?Y[%move(?X)] : − ?Y:block, ?Y[clear], ?X[clear], ?X[widerThen(?Y)],
                     btdelete{?Y[on ->?Z]}, btinsert{?Z[clear]},
                     btinsert{?Y[on ->?X]}, btdelete{?X[clear]}.

The sequential and concurent Transaction Logic implementations from University of Toronto and the Middle East University (Pinar Senkul, Fethi Altunyuva) develop a correct, but not efficient system for path execution.

**Conclusions and Future Work**

In this paper we studied the optimizations of Transaction Logic. We introduced two complementary methods of extending the inference system of transaction logic: tabling and incremental evaluation. The implementation of these methods and the performance of the trie based approach are still in progress and a future paper will talk about the data structures used to implement rule and state tries. The application of these algorithms promises great improvements in the applications of transaction logic dealing with systems with state, dynamic constraints on transaction execution and planning (STRIPS-like)

One particular effect of tabling is that tabled execution terminates more often than the execution based on depth first inference systems without tabling [VerbaetenEtAll01]. In our future work we want to study the termination of transaction logic programs with tabling. We also plan to study the effects of the negation, concurrent conjunction, possibility operators over the tabling for Transction Logic. We also want to study the development of a WAM that incorporates incremental + state/call tabling.

10